\documentclass[a4paper]{article}

\usepackage[pages=all, color=black, position={current page.south}, placement=bottom, scale=1, opacity=1, vshift=5mm]{background}
\SetBgContents{
	\tt 
}      % copyright

\usepackage[margin=1in]{geometry} % full-width

\usepackage{amsmath}
\usepackage{amsthm}
\usepackage{amssymb}

\usepackage[utf8]{inputenc}
\usepackage{hyperref}
\hypersetup{
	unicode,
	pdfauthor={Author One, Author Two, Author Three},
	pdftitle={A simple article template},
	pdfsubject={A simple article template},
	pdfkeywords={article, template, simple},
	pdfproducer={LaTeX},
	pdfcreator={pdflatex}
}

\usepackage[sort&compress,numbers,square]{natbib}

\theoremstyle{plain}

\theoremstyle{definition}

\usepackage{graphicx, color}
\graphicspath{{fig/}}

\usepackage{algorithm, algpseudocode} % use algorithm and algorithmicx for typesetting algorithms
\usepackage{mathrsfs} % for \mathscr command

\usepackage{lipsum}

\title{Developing a Self-Explanatory Transformer}
\author{Rasha Karakchi \and Ryan Karbowniczak}

\date{
	University of South Carolina\\ \texttt{karakchi@cec.sc.edu}\\%
	\texttt{karbowniczak@email.sc.edu}\\[2ex]%
}

\begin{document}
	\maketitle
	
	\begin{abstract}
		While IoT devices provide significant benefits, their rapid growth results in larger data volumes, increased complexity, and higher security risks. To manage these issues, techniques like encryption, compression, and mapping are used to process data efficiently and securely. General-purpose and AI platforms handle these tasks well, but mapping in natural language processing is often slowed by training times. This work explores a self-explanatory, training-free mapping transformer based on non-deterministic finite automata, designed for Field-Programmable Gate Arrays (FPGAs). Besides highlighting the advantages of  this proposed approach in providing real-time, cost-effective processing and dataset-loading, we also address the challenges and considerations for enhancing the design in future iterations.
	\end{abstract}

	%\tableofcontents
	
	\section{Introduction}
	The rapid increase in IoT devices, including sensors, cameras, and smart appliances, is leading to a significant surge in data generation \cite{mao2017survey}. Each device continuously produces data streams, resulting in an exponential rise in overall data volume. Edge computing addresses this challenge by processing data closer to its source, which helps reduce latency and bandwidth usage. However, this also requires large volumes of data to be processed, filtered, and analyzed in real-time, adding to the system's complexity. As data generation grows, so do the challenges of securing it. Edge computing introduces new points of vulnerability, necessitating robust security measures to protect data at the edge, during transmission, and within central data centers.
This increase in data has made data transformation a crucial aspect of modern computing. Techniques like data compression, filtration, encryption, and mapping are essential for ensuring faster, more reliable, and efficient data movement across systems. Data mapping, a key type of data transformation, is particularly important in natural language processing, where it involves converting or aligning textual data from one format or representation to another. This process is vital for tasks such as language translation, converting text into structured data, or aligning datasets from various sources. 

Automata-based transformation, a rule-based and training-free model, offers an efficient approach to data transformation and mapping by processing streamed input and generating equivalent output based on preloaded rules. General-purpose platforms have shown good results with deterministic automata-based transformations, where each input sequence produces only one output sequence \cite{nguyen2022gpu, nourian2022data}. However, some applications involve non-deterministic outputs like traveling salesman problem where multiple paths are explored, which are challenging to execute on general-purpose computing platforms like CPUs and GPUs due to the unpredictable memory accesses. Consequently, there is a shift towards domain-specific platforms like FPGAs and ASICs, which offer customization, on-chip memory usage, and parallel processing capabilities.

This poster presents a preliminary work on developing a data transformer using non-deterministic finite automata transduction techniques. Automata graph consists of states and edges carrying a combination of input symbol and output symbol. We design our accelerator as a two-dimensional array of processor elements (PEs) where each processor represents an edge, adapted from Micron's AP \cite{micron,reconfig}. Switch-based interconnection is designed to allow processor elements to communicate with and through neighbor PEs. We tested our design on programmable logic of Zync 104 FPGA and we measured the hardware resources and power consumed for the proposed approach.

	\section{Background}
	The Finite State Transducer (FST) is a finite state machine equipped with two memory tapes, similar to the input and output tapes of Turing machines. The processing unit acts as the state machine, moving between states based on input symbols and generating corresponding outputs. A Finite State Transducer (FST) is a 7-tuple (Q, $\sum$, $\delta$, $\omega$, $\Gamma$, q0, F) where Q set of states, $\sum$ set of symbols, $\delta$: Q x $\sum$ U $\epsilon$ $\rightarrow P(Q)$ the transition function, $\omega$: Q x $\sum$ U $\epsilon$ $Q \rightarrow {\Gamma}$   the output function (transduction), $q_0 \in Q$ the start state, and $F \subseteq Q$ a set of accepting states. 

In FST, data transformation is guided by an encoded relation, known as a rational or regular relation, which maps a set of input symbols to a set of output symbols. During each clock cycle, the automaton transitions based on (1) the current state and (2) whether the input symbol matches the edge label. When a match occurs and the automaton reaches a "report" state, it outputs a sequence.
\begin{figure}[h]
\centering
\includegraphics[width=0.64\linewidth]{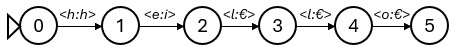}
  \caption{Example of Finite State Transducer}
  \label{fig:FST}
\end{figure}
\noindent
Figure \ref{fig:FST} illustrates an example of an FST with six states (where state 0 is the start state and state 5 is the report state) that recognizes the pattern "hello" and transforms it into "hi". Each transition is labeled with a \textit{Symbol1:Symbol2} pair, where \textit{Symbol1} is the input symbol triggering the transition, and \textit{Symbol2} is the output symbol generated. In this FST, each transition verifies the input symbol and, if it matches, produces the corresponding output symbol.
\noindent

\section{Proposed Approach}
Our design draws inspiration from NAPOLY \cite{napoly,karakchi2024scored,asap}, which is a non-deterministic finite automata overlay. It consists of an array of processor elements (PEs), each containing a $256x1-bit$ distributed on-chip RAM that stores the input symbol for the edge. Additionally, each PE is equipped with programmable switches that enable communication with neighboring PEs. The output function of the transformer is implemented using a separate on-chip distributed RAM.
\begin{figure}[h]
  \centering
\includegraphics[width=\linewidth]{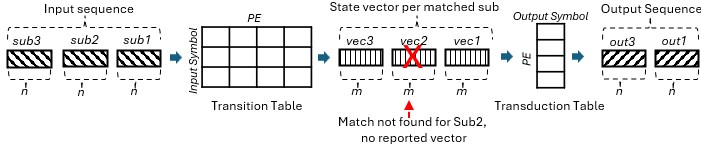}
  \caption{Transformer Framework}
  \label{framework}
 \end{figure}
 \noindent

Given the non-deterministic nature of the system, the path (i.e., the number of active states at one time) must be tracked while the input sequence is streamed to determine the corresponding output sequence. To manage this, we split the input sequence into fixed-length sub-sequences of size \textit{n}, as depicted in Figure \ref{framework}. For each sub-sequence, a vector is created to log the activated PE IDs. The maximum size of this vector is $O(m{^2})$, where \textit{m} represents the total number of PEs. If a match is detected within the array, the vector is flushed from the engine to the FIFO, and then to the transduction RAM, to generate the output sequence. After the vector is flushed, a new sub-sequence begins streaming into the array. If no match is found (i.e., no sequence is detected), the vector is discarded, and a new input sub-sequence begins streaming, as illustrated in Figure \ref{framework} for \textit{Sub2}.
\begin{figure}[h]
  \centering
\includegraphics[width=0.7\linewidth]{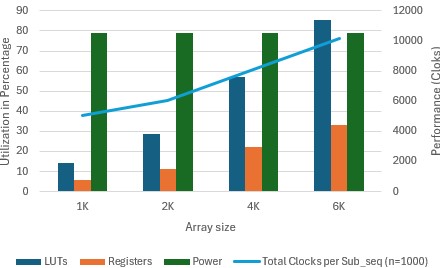}
  \caption{Preliminary Hardware Results}
  \label{results}
 \end{figure}
\noindent
In this design, we assume that each input symbol corresponds to exactly one output symbol. Consequently, the lengths of the output sub-sequences (\textit{out1, out3}) are equal to \textit{n}. This design is training-free and straightforward because new datasets are uploaded during reconfiguration, leveraging the on-chip memory to store the transition and output symbols. The execution time of the design is determined by the total time required to generate the output for a single sub-sequence.
The total execution time is calculated as follows:
$ Time\_to\_flush\_sub\_sequence (1\_clk x n) + transition\_time (2\_clk x n) + 
time\_to\_flush\_out\_vector (1\_clk) \\+ 
transduction\_time (1\_clk x m) + time\_to\_flush\_output\\ (1\_clk x n) $

\section{Preliminary Results and Conclusion}
We implemented the design on the programmable logic of the Zynq 104 FPGA. Figure \ref{results} illustrates the hardware resources used, power consumption, and memory utilization across various array sizes. The figure also demonstrates the increase in hardware resources (LUTs and Registers) relative to the array size, while power consumption remains at 79\% for all configurations since all arrays were run at the same maximum frequency of 500 MHz.

There are several limitations in the current design that we plan to address in future work. First, the PEs can only communicate with their immediate neighbors. While this reduces the number of required input/output wires and allows for larger array sizes, it also prevents direct communication between non-adjacent PEs, increasing the likelihood of congestion. Another limitation is the size and number of vectors generated for each sub-sequence to record the sequence track, which is constrained. In applications with many active states at once (i.e., multiple potential paths), tracking all paths becomes challenging which we aim to enhance for our future work. We also plan to explore streaming multiple input batches to improve performance.

%	\newpage
	\bibliography{main}

\end{document}